\documentstyle[epsfig]{ichep}
\begin{document}
\title{Electroweak theory at {\boldmath $Z^0$} and above $^{\dag}$}
\author{M.\ Vysotsky$^{*}$}
\affil{ITEP, B.\ Cheremushkinskaya 25, Moscow 117259, Russia}
\abstract{We present a simple method for considering radiative corrections.
Precision data are well described by an adequate Born approximation and
only the first evidence for deviations from the tree level formulas and the
presence of the loop effects emerge.}
\twocolumn[\maketitle]
\fnm{7}{E-mail: VYSOTSKY@VXITEP.ITEP.MSK.SU}
\fnm{6}{Plenary talk given at 27th International Conference on High Energy
Physics, Glasgow, 20-27th July 1994}
\section{Introduction}
In my talk I will concentrate on electroweak radiative corrections (EWRC) in
the Minimal Standard Model (MSM) [1-2]. To discuss radiative
corrections one should start from the Born approximation to the theory. When
dealing with the Born approximation, it is convenient to define the parameters
of the theory at the characteristic energy scale (or momentum transfer) of the
problem. Thus we avoid large loop corrections. This scale in electroweak
theory is determined by the masses of the intermediate vector bosons. To
start with, I shall introduce this natural Born approximation and then I
will show that the experimental data on $Z$ decays and $m_W$ measurements
agree rather well with it. However, some of the latest data as shown in the
talk by D.\ Schaile \cite{100} differ from the Born approximation at the
level of two standard deviations. Then I will take into account EWRC. The
bound on $m_{top}$ will arise from by demanding their smallness. Taking EWRC
into account, we will describe all the data within one standard deviation
with the exception of $\Gamma_{Z\to b\bar{b}}$. Then I will discuss how
electroweak theory will be tested in the future. This talk is based on the
approach to the calculation of EWRC developed in papers [4-11].
\section{Born approximation to MSM}
To predict something in the electroweak theory one should specify the
numerical values of its parameters: $SU(2)$ and $U(1)$ gauge coupling
constants $g$ and $g'$, Higgs vacuum expectation value  $\eta$, Higgs
quartic coupling $\lambda$, masses of quarks and leptons and quark mixing
angles. Let us discuss the first three parameters. To determine them one
takes the three best measured quantities: the muon decay constant $G_{\mu}
=1.16639(2)\cdot 10^{-5}$ GeV$^{-2}$, $m_Z = 91.1888(44)$ GeV, measured at
LEP \cite{9}, and the value of the running electromagnetic coupling at
$q^2 = m^2_Z$, $\bar{\alpha}\equiv \alpha(m_Z)=[128.87(12)]^{-1}$ \cite{10}.
The relevant scale for $G_{\mu}$ is not the momentum transfer which is of
the order of the muon mass, but $m_W$. This is clear from the form of
$W$-boson propagator:
\begin{equation}
G_{W}^{-1}=q^2 -m_W^2 -\Pi_W(q^2) \;\; .
\label{1}
\end{equation}
The logarithmic dependence of the $W$-boson polarization operator on $q^2$
is power suppressed relative to $m^2_W$. So only $\alpha$ runs below $m^2_W$.
We should renormalize it from the low energy fine structure value to the
value at $m^2_Z$.

The value of $\bar{\alpha}$ is described by the following well-known formula:
\begin{equation}
\bar{\alpha}=\frac{\alpha}{1-\delta\alpha} \;\; .
\label{2}
\end{equation}
Here $\delta\alpha$ with the help of a dispersion relation is expressed
through the integral of the cross section for $e^+e^-$-annihilation:
\begin{equation}
\delta\alpha=\frac{m_Z^2}{4\pi^2\alpha}
\int^{\infty}_{th}\frac{\sigma_{e^+e^-}(s)ds}{m_Z^2 -s} \;\; .
\label{3}
\end{equation}
The charged lepton contributions to $\delta\alpha$ are easily calculated:
\begin{equation}
(\delta\alpha)_l =\frac{\alpha}{3\pi}[\sum_{l}(\ln\frac{m_Z^2}{m_l^2}-
\frac{5}{3})] = 0.0314 \;\; .
\label{4}
\end{equation}

The result for the hadrons contribution which follows from integrating the
experimental data is \cite{10}:
\begin{equation}
(\delta\alpha)_h = 0.0282(9) \;\; .
\label{5}
\end{equation}

The main source of uncertainty in $(\delta\alpha)_h$ is the systematic
uncertainty in the cross section for $e^+e^-$-annihilation into hadrons
between the $\rho$- and $\Psi$-mesons. There are two ways to improve the
accuracy of $\delta\alpha$: to measure $\sigma_{e^+e^-\to hadrons}$ with
better accuracy and/or to improve the experimental accuracy in the
$(g-2)_{muon}$ measurement. As the hadronic contribution to $(g-2)_{muon}$
is proportional to the integral of $\sigma_{e^+e^-}(s)/s$, it is sensitive
to the low energy hadronic states.

It is interesting that the result for $(\delta\alpha)_h$ obtained from
integrating the experimental data can be reproduced in the following simple
model: the lowest lying resonance in each flavor channel as an infrared
cutoff plus perturbative QCD continuum \cite{11}. The main question in this
model:  what is the value of the $(\delta\alpha)_h$ uncertainty? It appears
that the uncertainties of the vector meson's width into $e^+e^-$ and of
$\alpha_s(m_Z)$ lead to a very small variation of $(\delta\alpha)_h$. To
get the uncertainty $\pm 0.0009$, we need a 30\% variation of perturbative
$\sigma_{e^+e^-\to hadrons}$ in the domain 1 GeV $<\;E<\;$ 2 GeV.

In a QCD sum rules inspired model \cite{12} the following result was obtained:
\begin{eqnarray}
(\delta\alpha)_{S.R.}&=&0.0275(2) \;\; , \nonumber \\
(\bar{\alpha})_{S.R.}&=&[128.96(3)]^{-1} \; ,
\label{6}
\end{eqnarray}
where the uncertainty is very small. Below I will use an $\bar{\alpha}$ value
which follows from the integration of the experimental data \cite{10}:
\begin{equation}
\bar{\alpha}=[128.87(12)]^{-1} \;\; .
\label{7}
\end{equation}

Precision measurement of $\sigma_{e^+e^-\to hadrons}$ below 1.4 GeV is now
under way in Novosibirsk \cite{13}.

To present Born approximation formulas for different observables we should
introduce an electroweak mixing angle. Among several existing definitions we
use one directly connected with the most precisely measured quantities
\cite{14}:
\begin{eqnarray} \sin^2\theta\cos^2\theta&\equiv&
c^2s^2=\frac{\pi\bar{\alpha}}{\sqrt{2}G_{\mu}m^2_Z} \;\; ,\nonumber \\
\sin^2\theta &=& 0.2312(3)
\label{8}
\end{eqnarray}
where the uncertainty in $\sin^2\theta$ arises from that in $\bar{\alpha}$.
I will start from the observables which depend on strong interactions only
through the higher loops (``gluon-free" observables): $W$-boson mass $m_W$,
$Z$-boson width into charged leptons $\Gamma_l$ and forward - backward
asymmetries in $Z\to l^+l^-$ decay $A^l_{FB}$. It is convenient to consider
the axial and vector coupling constants of the $Z$ boson with the charged
leptons which are defined from $\Gamma_l$ and $A^l_{FB}$:
\begin{eqnarray}
\Gamma_l &=&4(1+\frac{3\alpha}{4\pi})(g^2_A+g^2_V)\Gamma_0 \;\; ,\nonumber \\
\Gamma_0 &=&\frac{\sqrt{2}G_{\mu}m^3_Z}{48\pi}=82.95 MeV \;\; ; \nonumber \\
A^l_{FB} &=& \frac{3(g_V/g_A)^2}{[1+(g_V/g_A)^2]^2} \;\; .
\label{9}
\end{eqnarray}
Since $g_V$ turned out to be very small, the quark forward - backward
asymmetries define its value as well.

In table 1 the Born approximation predictions for the gluon-free observables
are compared with the experimental data presented at the Marseille (1993)
Europhysics conference \cite{112} and this conference [12,18,19].
While a year ago the Born description was well within one standard
deviation from experimental data \cite{5}, now deviations at the level of
$2\sigma$ for $\Gamma_l$ and $1.5 \sigma$ for $m_W$ occur. So the first
evidence for electroweak radiative
corrections from the latest experimental data emerges \cite{8}. If we
substitute instead of $\bar{\alpha}$ the value of the fine structure
constant in the definition of electroweak mixing angle then we will obtain
$\sin^2\theta_W = 0.212$ and $g_V/g_A = 0.152$ which is more than 50
standard deviations from the LEP+SLC result. This is the accuracy with
which the running of $\alpha$ electromagnetic is confirmed.
\begin{table*}
\Table{|c|c|c|c|}{
 &BORN & MARSEILLE (1993) & GLASGOW (1994)\\ \hline
$(m_W/m_Z)_{p\bar{p}}$ & 0.8768(2) & 0.8798(28) &
 0.8798(20) \\ \hline
  &   &    &   LEP: 0.0716(16) \\
 $(g_V/g_A)^{all~~asymm.}_{LEP, SLC}$ & 0.0753(12) & LEP: 0.0712(28) & SLC:
 0.0824(40) \\ &   &    &  LEP+SLC: 0.0731(15) \\ \hline
$\Gamma_l(MeV)_{LEP}$ & 83.57(2) & 83.79(28) & 83.96(18) \\ \hline}
\caption{
Comparison of the Born approximation predictions for gluon-free observables
with experimental data. Errors on the Born values come from that in
$\bar{\alpha}$.}
\end{table*}

Now come strong interaction sensitive observables. Here we take the EW Born
approximation improved by strong interaction perturbation theory and
this is the formula for $\Gamma_{Z\to hadrons} \equiv\Gamma_h$:
\begin{eqnarray}
\fl~~~~~\Gamma_h^B=3\Gamma_0\{3[1+(1-\frac{4}{3}s^2)^2] +
2[1+(1-\frac{8}{3} s^2)^2]\}\cr
[1+\frac{\bar{\alpha}_s}{\pi} + 1.4(\frac{\bar{\alpha}_s}{\pi})^2 -
13(\frac{\bar{\alpha}_s}{\pi})^3]  \; .
\label{10}
\end{eqnarray}
Substituting $\bar{\alpha}_s\equiv\alpha_s(m_Z)_{\overline{MS}}= 0.125 \pm
0.005$ (see below), we get:
\begin{equation}
\Gamma^B_h =
1748(3) MeV \; , \;\; \Gamma^B_Z = 2496(3) MeV \;\; ,
\label{11}
\end{equation}
which is within one standard deviation of the experimental numbers \cite{9}:
\begin{eqnarray}
\Gamma^{exp}_h&=&1745.9\pm 4.0 MeV \nonumber \\
\Gamma^{exp}_Z&=&2497.4\pm 2.7\pm 2.7 MeV
\label{12}
\end{eqnarray}
Even for $\Gamma_{Z\to b\bar{b}}$ the Born approximation describes
experimental data within one standard deviation:
\begin{eqnarray}
\Gamma^B_{Z\to b\bar{b}}
&=& 3\Gamma_0 G[1+(1-\frac{4}{3}s^2)^2] = \nonumber \\
&=&383.2(6) MeV \; ,
\\
\Gamma^{exp}_{Z\to b\bar{b}}&=& 384.5\pm 3.5 MeV \mbox{\cite{18}} \;\; .
\nonumber
\label{13}
\end{eqnarray}
\section{EWRC}
Taking into account radiative corrections, we obtain the following
formulas for gluon-free observables:
\begin{eqnarray}
m_W/m_Z &=& c+\frac{3c\bar{\alpha}}{32\pi s^2(c^2-s^2)}V_m(t,h) \;\; ,
\nonumber \\
g_A &=& -1/2-\frac{3\bar{\alpha}}{64\pi s^2c^2}V_A(t,h) \;\; ,\\
g_V/g_A &=& 1-4s^2 +\frac{3\bar{\alpha}}{4\pi(c^2 -s^2)}V_R(t,h) \;\; ,
\nonumber
\label{14}
\end{eqnarray}
where $t\equiv(m_t/m_Z)^2$, $h\equiv(m_H/m_Z)^2$. The coefficients in front
of $V_i$ are fixed by the assymptotic behavior for a very heavy top:
$V_i(t\to\infty)\to t$. $V_i$ should not be large in order not to spoil the
good description of the experimental data by the Born formulas. From
experimental data we get:
\begin{eqnarray}
V^{exp}_m &=& 1.82\pm 1.28 \;\; , \nonumber \\
V^{exp}_A &=& 2.00\pm 0.77 \;\; , \\
V^{exp}_R &=& -0.63\pm 0.53 \;\; .\nonumber
\label{15}
\end{eqnarray}
The $V_i$ dependence on $t$ appeares to be rather similar and they equal zero
around $m_t = 150$ GeV. This is the origin for the prediction of an $m_{top}$
value around 170 GeV from the precise measurements.

For $Z$-boson decay into quarks the analogous formulas hold for the axial
and vector coupling constants:
\begin{eqnarray}
g_{Af}&=&T_{3f}[1+\frac{3\bar{\alpha}}{32\pi s^2c^2}V_{Af}(t,h)] \;\;
,\nonumber \\
g_{Vf}/g_{Af}&=&1-4|Q_f|s^2+  \\
&+&\frac{3\bar{\alpha}|Q_f|}{4\pi(c^2-s^2)}V_{Rf}(t,h) \;\; , \nonumber
\label{16}
\end{eqnarray}
where $f$ indicates the quark flavor.

In the case of light $u-$, $d-$, $s-$ and $c-$ quarks the only difference
with leptonic $V_A$ and $V_R$ are in small constants $c^f_{A,R}$ \cite{4}:
\begin{equation}
u,d,s,c: V^f_{A,R} = V_{A,R} + c^f_{A,R} \;\; .
\label{17}
\end{equation}
There are 3 additional terms in the case of $Z\to b\bar{b}$ decay:
\begin{eqnarray}
\Gamma_{b\bar{b}}&=&\Gamma_{d\bar{d}}-\Gamma_0[\frac{\bar{\alpha}}{\pi}
\phi(t) +(\frac{\bar{\alpha}_s}{\pi})^2 I(t) + \nonumber \\
&+& 3\times 6(\bar{m}_b/m_Z)^2]\equiv \Gamma_{d\bar{d}}-\varepsilon\Gamma_0
\;\; ,
\label{18}
\end{eqnarray}
where $\bar{m}_b\equiv m_b(m_Z)$. The first term originated from the
vertex with a virtual top [21-23], the second from an imaginary part
of the 3 loop insertion into the $Z$-boson propagator with a 2 gluon
intermediate state [24-25] and the reason for the third is the nonzero
$b$-quark mass. As it was demonstrated in \cite{24} the running $b$-quark
mass at $q^2 = m_Z^2$ enters when the gluon corrections are taken into
account.

{}From the ratio $\Gamma_b/\Gamma_h$ all nonspecific to $b$-quarks EWRC
almost cancel out, as well as the universal $\bar{\alpha}_s$ corrections.
As a result we obtain a tree level formula improved by the specific terms:
\begin{eqnarray}
\frac{\Gamma_b}{\Gamma_h}&\approx&\frac{1+(1-\frac{4}{3}s^2)^2}
{5+2(1-\frac{8}{3}s^2)+3(1-\frac{4}{3}s^2)^2}\times \nonumber \\
&\times& [1-\frac{\varepsilon}{3}(\frac{1}{1+(1-\frac{4}{3}s^2)^2} -
\nonumber \\
&-&\frac{1}{5+2(1-\frac{8}{3}s^2)^2 +3(1-\frac{4}{3}s^2)^2})] =
\nonumber \\
&=& 0.2197\{1-0.176[\frac{\bar{\alpha}}{\pi}\phi(t)+ \nonumber \\
&+&(\frac{\bar{\alpha}_s}{\pi})^2 I(t)+0.019]\} \; ,
\label{19}
\end{eqnarray}
where we substitute $\bar{m}_b = 3.0$ GeV. Expanding $\phi$ and $I$ around
$m_t = 150$ GeV we get:
\begin{eqnarray}
\frac{\bar{\alpha}}{\pi}\phi(t) &=& 0.052[1+1.4\frac{m_t - 150}{m_Z} + ... ]
\nonumber \\ \\
(\frac{\bar{\alpha}_s}{\pi})^2 I(t) &=& 0.018[1+0.3\frac{m_t - 150}{m_Z} +
\ldots] \nonumber
\label{20}
\end{eqnarray}
where we use $\bar{\alpha}_s=0.12$. So in the expression for $\varepsilon$
the first term dominates while the other two are almost equal. Substituting
$m_t=175$ GeV we obtain $(\Gamma_b/\Gamma_h)=0.2154$ which is very close to
the result which takes into account the neglected EWRC:
\begin{equation}
(\Gamma_b/\Gamma_h)_{theor} =
0.2161(4)^{-6\leftarrow m_H = 1000 GeV}_{+6 \leftarrow m_H = 60 GeV} \;\; ,
\label{21}
\end{equation}
where the central value corresponds to $m_H=300$ GeV, $m_t=175$ GeV and
the shift due to a $\pm 10$ GeV $m_t$ variation is indicated in brackets.
Comparing with the experimental result \cite{18}:
\begin{equation}
(\Gamma_b/\Gamma_h)_{exp} = 0.2202(20)
\label{22}
\end{equation}
we see that LEP data show a $2\sigma$ deviation from the MSM result.

If $\Gamma_{b\bar{b}}$ really contradicts the MSM, what kind of new physics
can describe this? New physics contributions to the vector boson polarization
operators modify the functions $V_i$, which cancel out
from the ratio $\Gamma_b/\Gamma_h$. But SUSY contributes not only through
$V_i$ modifications. Two new types of vertex diagrams with an intermediate top
and stop appear: with $(t,H^+)$ exchange and with $(\tilde{t}, \tilde{W}^+)$
exchange. As it was noted in \cite{25} the diagram with $(\tilde{t},
\tilde{W}^+)$ exchange enlarge $\Gamma_{b\bar{b}}/\Gamma_h$, so SUSY may be a
solution if the discrepancy with the MSM is confirmed. According to \cite{26}
in order for SUSY to resolve the discrepancy with experiment superpartners
should be rather light: $m_{\tilde{t}},~m_{\tilde{W}^{\pm}}<100$ GeV.

The LEP I Precision Calculations Working Group was organized in CERN with the
following aims: to update the existing electroweak libraries; to check the
reliability of independent calculations; and to estimate the uncertainties of
theoretical predictions. There were three meetings of this group in 1994
and the results obtained are summarized in the contribution by G.\ Passarino
\cite{27}. The program LEPTOP, which was written by A.N.\ Rozanov and which
was used by him to obtain the numerical results which I cite below, was
updated in the framework of Precision Calculations Working Group.

A number of different higher-order corrections were calculated during the
last two years. The imaginary parts of the diagrams shown in figures 1-4
contribute to the Z width. In the figure captions the calculated terms are
designated.
\begin{figure}
\centerline{\epsfig{file=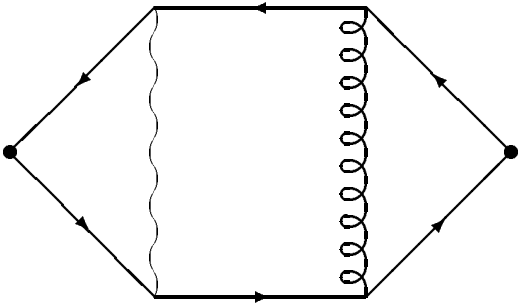}}
\caption{The contribution of these types of diagram to $\Gamma_Z$ were
calculated in [30]. Here a solid line is the quark propagator, a wavy line
--- the photon  propagator, a curly line --- the gluon propagator. The
correction to $\Gamma_Z$ is of order $\bar{\alpha}\bar{\alpha}_s$.}
\end{figure}
\begin{figure}
\centerline{\epsfig{file=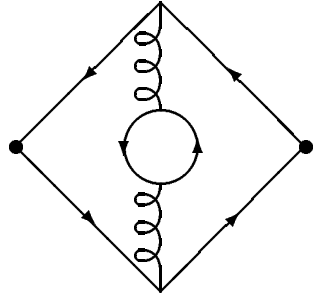}}
\caption{
The correction to $\Gamma_Z$ of order $\bar{\alpha}^2_s \frac{m^2_Z}{m^2_t}$
originated from these types of diagrams where the top quark in the inner
loop was calculated in [31].}
\end{figure}

\begin{figure}
\centerline{\epsfig{file=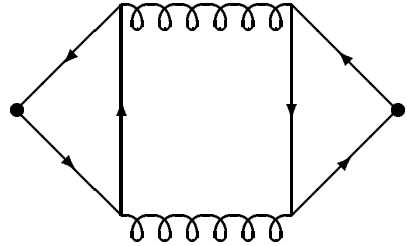}}
\caption{
The correction to $\Gamma_Z$ of order $\bar{\alpha}^2_s \frac{m^2_b}{m^2_Z}
f(t)$ was calculated in [32].}
\end{figure}
\begin{figure}
\centerline{\epsfig{file=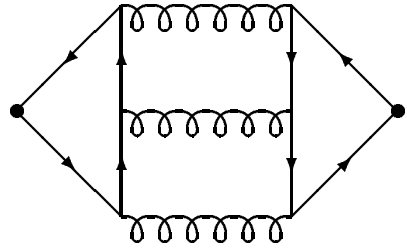}}
\caption{
The $\bar{\alpha}^3_s\chi(t)$ correction to $\Gamma_Z$ was calculated in
[33,34].}
\end{figure}

For the difference of $Z$- and $W$-bosons polarization operators at zero
momentum transfer the correction of order $\alpha_w \alpha^2_s t$ was
calculated recently \cite{33}. The final result looks like:
\begin{leqnarray}
\label{23}
\fl\frac{\Pi_Z(0)}{m^2_Z} - \frac{\Pi_W(0)}{m^2_W} \sim \cr
 \alpha_w t[1-0.9
\alpha_s(m_t) - (2.2 - 0.2 n_f) \alpha^2_s(m_t)]\;,
\end{leqnarray}
where $n_f=5$ is the number of quark flavors light in comparison with $m_t$.

The leading term in the intermediate vector bozon polarization operators
difference $\sim \alpha_w t$  was calculated long ago \cite{34}. The strong
interaction correction $\sim \alpha_w \alpha_s(m_t)t$ was calculated in
\cite{35}, where the answer for arbitrary $m_t$ was also presented. The
correction $\sim\alpha^2_w t^2$ due to the diagrams with Higgs exchange was
calculated in \cite{36}. The correction $\sim \alpha_w\alpha^2_s(m^2_t)t$ due
to the diagram shown in figure 3 was calculated in \cite{37}, the contribution
of the same order of diagram with the insertions into a gluon propagator (the
same topology as shown in figure 2) was estimated in \cite{38} and the final
answer (eq. 23) was  derived in \cite{33}. Taking all these results we obtain
the following prediction for the $W$-boson mass:
\begin{leqnarray}
\label{eq24}
\fl m_W(MeV) = \stackrel{tree}{79958(18)} + \stackrel{\alpha_w}{445} -
\stackrel{\alpha_w\alpha_s}{65} - \stackrel{\alpha^2_w}{18} - \cr
-\stackrel{\alpha_w\alpha^2_s}{8} = 80312\;,
\label{24}
\end{leqnarray}
where we use $m_t=175$ GeV, $m_H=300$ GeV, $\bar{\alpha}_s=0.125$. The error
in tree level result is due to that in $\bar{\alpha}$. You see that a
theoretical error in $m_W$ due to still uncalculated terms should be very
small.
\section{Global Fit}
The experimental data from LEP, hadron colliders and SLC which we use in
the global fit are shown in table 2. From each of these data for fixed $m_H$
and $\bar{\alpha}_s$ values a prediction for top quark mass can be obtained,
see figure 5.
\begin{table*}
\Table{|ccccc|}{
$\Gamma_Z$ (GeV) & $\sigma^0_h$ (nb) & $R_l$ & $A^{0, l}_{F B}$ & $A_{\tau}$ \\
2.4974(38)      & 41.49(12)  & 20.795(40) & 0.0170(16) & 0.143(10) \\ \hline
$A_l$  & $R_b$  &  $R_c$ & $A^{0,b}_{FB}$ & $A^{0,c}_{FB}$ \\
0.135(11)  & 0.2202(20)  & 0.1583(98)  & 0.0967(38)  & 0.0760(91) \\ \hline
$\sin^2 \theta^{lept}_{eff} \rm{from} <Q_{FB}>$ & $m_W$ &
$1-m^2_W/m^2_Z(\nu N)$ & $\sin^2 \theta^{lept}_{eff} \rm{from}~ A_{LR}$ &
\\ 0.2320(16) & 80.23(18) & 0.2256(47) & 0.2294(10) & \\ }
\caption{Input data for global fit [3]. All numbers except the last three are
from LEP.}
\end{table*}
\begin{figure*}
\centerline{\epsfig{file=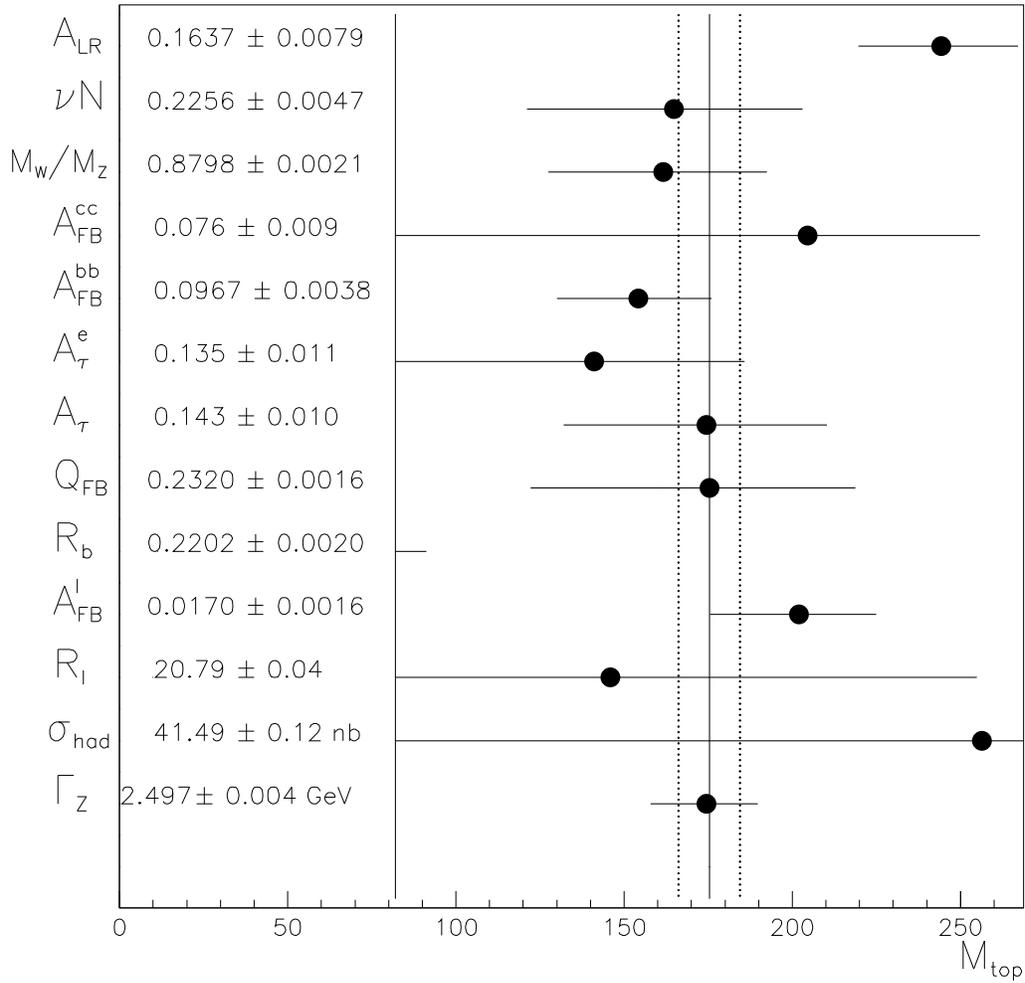,height=15cm}}
\caption{The values of $m_{top}$ from different precision measurements for
fixed $m_H = 300$ GeV and $\bar{\alpha}_s = 0.125$. Vertical solid line
corresponds to average value $m_t = 176$ GeV, while dotted lines to $\pm9$
GeV experimental uncertainty.}
\end{figure*}

{}From the whole data set we get:
\begin{equation}
m_t = 176 \pm 9^{+19+6}_{-21-6} GeV
\label{25}
\end{equation}
\begin{eqnarray}
\bar{\alpha}_s = 0.125 \pm 0.005 \pm 0.002\\
\nonumber
\chi^2/d.o.f. = 15/12\;,
\label{26}
\end{eqnarray}
where the central values correspond to $m_H=300$ GeV, the first error is
experimental, the second corresponds to $m_H=1000$ GeV $(+)$ and $m_H=60$
GeV $(-)$ and the third error in $m_t$ is due to the $\bar{\alpha}$
uncertainty: $(+)$ for $\bar{\alpha}=(128.75)^{-1}$ and $(-)$ for
$\bar{\alpha}=(128.99)^{-1}$.

The recently calculated correction of order $\bar{\alpha}_w\alpha^2_s$
to the vector boson polarization operators \cite{33} shift the central
value of $m_t$ in the following way:
\begin{equation}
174.5 GeV
\stackrel{\alpha_w\alpha^2_s}{\rightarrow} 176.0 GeV\;.  \label{27}
\end{equation}
This $1.5$ GeV shift demonstrates the present day theoretical accuracy in
the $m_t$ calculation from the precision data set. At this conference the
CDF collaboration reported evidence for $t$-quark production at the Fermilab
collider \cite{40}. Their value for the top mass is:
\begin{equation}
m^{CDF}_t = 174 \pm 10^{+13}_{-12} GeV
\label{28}
\end{equation}
Figure 6 shows the $\chi^2$ levels for a $(m_H\;, m_t)$ plane from the
global fit for the fixed value $\bar{\alpha}_s = 0.125$. As is evident
from this plot while stringent bound on $m_{top}$ emerge from the precise
data, no conclusive evidence for an $m_H$ upper bound beyond $1$ TeV exists.
So $m_H$ now is bounded from below by the LEP direct search bound $m_H>60$
GeV and from above by unitarity arguments: $m_H < 1$ TeV. Even if we take
into account the CDF number (28) figure 6 does not change qualitatively
and still no bound on $m_H$ emerges (see figure 7). The reason for this
misfortune, from the point of view of bounding the Higgs mass, is the
coincidence of the $m_t$ central value from the global fit (25) and
the CDF value (28). However, if in the future we have some luck and the top
appears to be light, than even the present day accuracy of precision data will
be enough to bound $m_{Higgs}$. For example with $m_t=145\pm5$ GeV from the
hadron collider $m_H<160$ GeV at 95\% (see figure 8)$^{\dag}$ \fnm{1}{This
remark is due to L.B.\ Okun.}.
\section{Above the {\boldmath $Z$}$^{\ddag}$}
\fnm{2}{For more detailed discussion of material covered in this section see
[42].}
A next step in the investigation of electroweak theory will occur after LEP II
starts to operate.

One of the main aims of LEP II is the measurement of the $W$-boson mass with an
accuracy better than $50$ MeV. The present day experimental number is:
\begin{equation}
m^{p\bar{p}}_W = 80.230 \pm 0.180 GeV\;.
\label{29}
\end{equation}
The theoretical prediction for $m_t=175$ GeV, $m_H=300$ GeV and
$\bar{\alpha}_s=0.125$ is:
\begin{equation}
m^{th}_W = 80.312 \pm 0.060^{+0.0}_{-0.018}\; GeV
\label{30}
\end{equation}
where the first error corresponds to $\pm 10$ GeV $m_t$ variation, while
the second corresponds to $m_H = 1000$ GeV $(+)$ and $m_H = 60$ GeV $(-)$.
When shifting $m_H$ we change $m_t$ to its global fit value and this is
the reason for small dependence of $m_W^{th}$ on $m_H$. If we take $m_t =
175$ GeV, then $\Delta M_{W}=-100$ MeV for $m_H=1000$ GeV and $\Delta
M_W=+100$ MeV for $m_H=60$ GeV. So a measurement of $m_W$ with high
accuracy will provide us with a new check of the minimal standard model,
will improve the bound on $m_t$ and will allow a possibility to bound
$m_H$ if $m_t$ is measured with accuracy better than $10$ GeV.

Feynman diagrams for pair of $W$-bosons production are shown on figure 9. To
determine value of $m_W$ one should reconstruct the kinematics of the event.
For this purpose the initial energy should be known and the error in its
determination will transfer into the error in $m_W\;,\;\Delta M_W\sim\Delta
E_{initial}=E_0-<E_{\gamma}>\;$, where $<E_{\gamma}>$ designates energy
radiated from the initial state. The following estimate of the radiated
energy takes place: $<E_{\gamma}>=1\div2$ GeV depending on the total $e^+e^-$
energy $\sqrt{s}=176\div190$ GeV. So the accuracy of the $<E_{\gamma}>$
calculation should be of the order of 1\%.
\begin{figure*}[p!]
\centerline{\epsfig{file=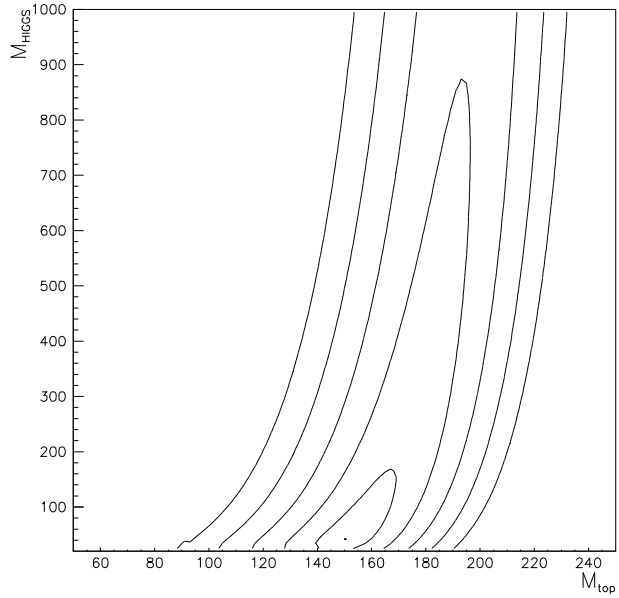,height=9cm}}
\caption{Levels of constant $\chi^2 = \chi^2_{min} + n^2\;, n = 1,2,3, ...$
from a global fit. The value of $\bar{\alpha}_s$ is fixed, $\bar{\alpha}_s =
0.125$.}
\end{figure*}
\begin{figure*}[p!]
\centerline{\epsfig{file=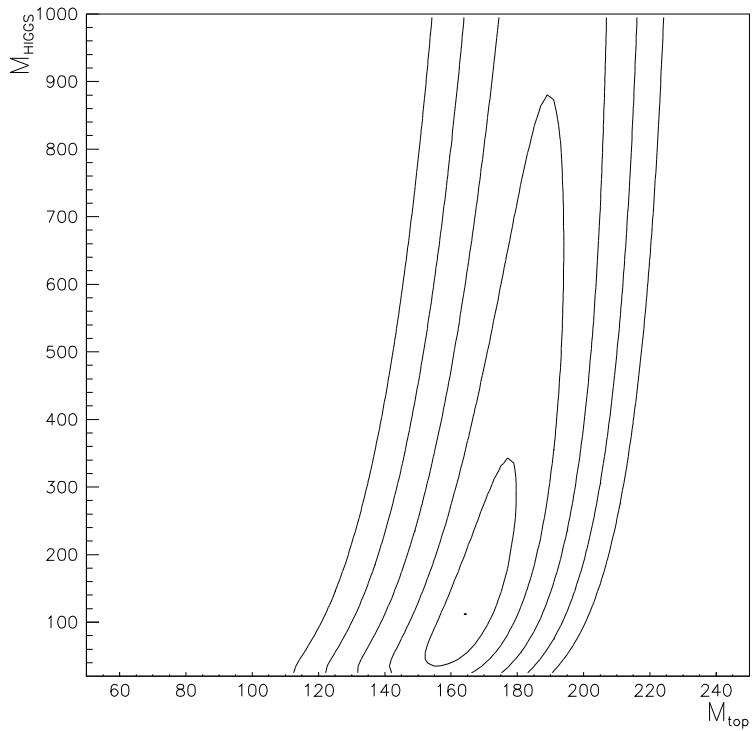,height=11cm}}
\caption{The same as figure 6 with the addition of the value of the
top quark mass from the CDF ``evidence": $m_t^{CDF}=174\pm 16$ GeV.}
\end{figure*}
\begin{figure*}[t!]
\centerline{\epsfig{file=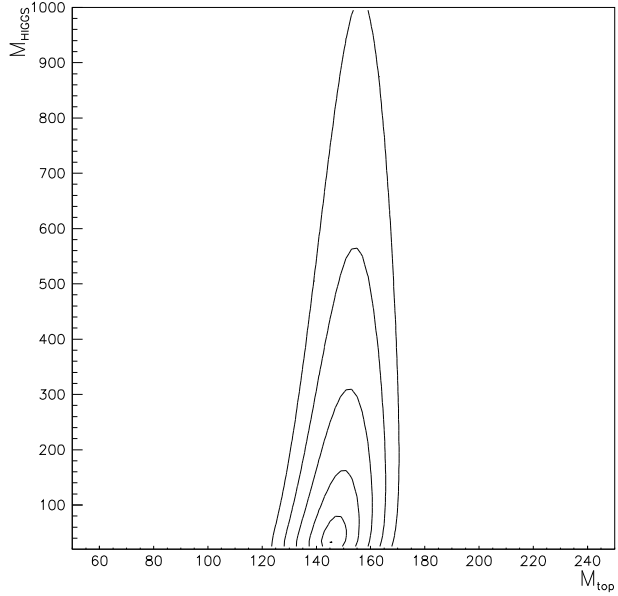,height=9 cm}}
\caption{The same as figure 6 with the addition of an imaginary result for the
value of the top quark mass from future measurements: ``$m_t=145\pm 5$ GeV".}
\end{figure*}
\begin{figure*}[t!]
\centerline{\epsfig{file=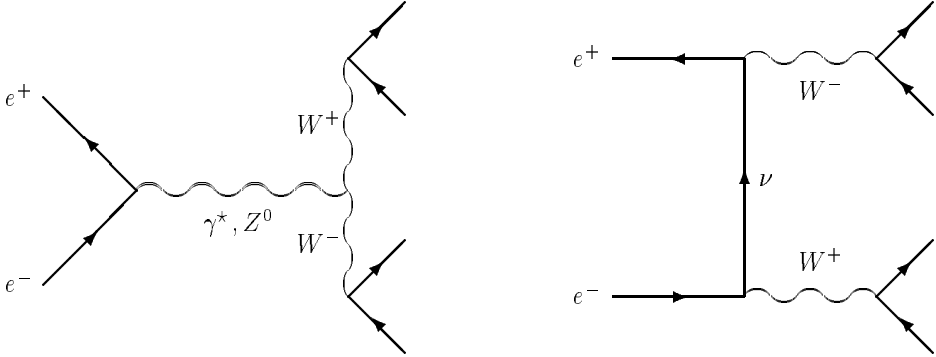}}
\caption{``Crayfish" and ``Crab" diagrams responsible for $W$-boson pair
production
in $e^+e^-$-annihilation.}
\end{figure*}
$<E_{\gamma}>$ is determined by the
following equation:
\begin{equation}
<E_{\gamma} >=\frac{1}{\sigma_T}\int\omega\frac{d\sigma}{d\omega}d\omega\;,
\label{31}
\end{equation}
where $\sigma_T$ is the inclusive cross section for $W^+W^-$ production,
while $d\sigma/d \omega$ is the differential cross section for $W^+W^-\gamma$
production. That is why tree level
formulas are not enough and radiative corrections (especially enhanced) should
be accounted for. Let me remind you that in the case of resonant $Z$ production
in $e^+e^-$ annihilation the tree level cross section is strongly damped by a
Sudakov dilogarithmic formfactor:
\begin{equation}
\sigma = \sigma_0 \exp \left\{-\frac{\alpha}{\pi} \ln (\frac{m^2_Z}{m^2_e})
\ln (\frac{m^2_Z}{\Gamma^2_Z})\right\} \;.
\label{32}
\end{equation}
For $W^+W^-$ production the second log is absent, however logarithmically
enhanced terms $\sim [\alpha \ln (\frac{s}{m^2_e})]^n$ due to radiation of
photons from the initial state (ISR) exist. The total cross section of ISR
is not gauge invariant, since unlike the case of $Z$ production the flow of
charge in the initial state is not continuous because of the ``crab" diagramm.
Nevertheless in paper \cite{42} using the so-called current spliting technique
a gauge-invariant definition of ISR is given.

The present day situation with the calculation of $W^+W^-$ production can
be summarized in the following way: tree level cross sections as well as
enhanced radiative corrections are calculated. Calculations of non-enhanced
radiative corrections are in progress. For a comprehensive review see
\cite{43}.

Another problem for LEP II is to study the  triple vector bozon vertices
$\gamma W^+W^-$ and $Z W^+W^-$. In the Standard Model these vertices are
determined by the non-abelian SU(2) coupling $g$. Present day Fermilab hadron
collider bounds on them are of the order of unity \cite{44}. According to
\cite{45} the accuracy in the measurement of the MSM couplings as well as
bounding anomalous triple vector boson couplings at LEP II will be of the
order of 0.1, while at the Next Linear $e^+e^-$ Collider with c.m. energy
500 GeV it would be 0.01 and for a c.m.\ energy 1000 GeV an accuracy up to
0.001 can be reached. This will allow us to study the triple vertices at the
level of MSM radiative corrections.
\section{Conclusions}

1. Accuracy in measurements of a number of electroweak observables now
reaches $2\cdot 10^{-3}$ which allows the data to start to be sensitive
to electroweak radiative corrections.

2. The set of precision data produces in the framework of MSM a very
stringent bound on $m_{top}$~, for further progress in bounding $m_{top}$
it is important to determine the value of $\bar{\alpha}\equiv\alpha(m_Z)$
with better accuracy.

3. $t$-quark discovery at the Tevatron (or an improved lower bound on its
mass) together with future improvements in LEP I and SLC data on $Z$-boson
parameters will allow us to bound the Higgs boson mass.

4. The most intriguing expectation from precision measurements is a deviation
from the MSM predictions, which will signal the presence of physics beyond the
standard model. May be we already see such a deviation in $\Gamma_{b\bar{b}}$.

I am grateful to my coauthors on papers [4-11] from which my
understanding of the subject of EWRC emerge and especially to V.A.\ Novikov
and L.B.\ Okun, to A.N.\ Rozanov who kindly did the computer calculations used
by me, to I.G.\ Knowles for assisting with the figures, to D.\ Schaile for
making
the latest experimental data available, to Jose
Valle for hospitality at the University of Valencia where this talk was
prepared and to ISF grant MRW000 and Russian Foundationn for Fundamental
Research grant 93-02-14431 for partial financial support.

\section*{Questions}
{\it H.E.\ Haber, UCSC:} \\
Concerning the global LEP fit and including the CDF ``measurement" of
$m_t$, the value of $m_H$ with minimum $\chi^2$ that you obtain seems
somewhat higher than the result shown earlier by Schaile (these
proceedings). Can you explain the discrepancy, or is the difference
within the uncertainties of the two analyses?
\vspace{5mm} \\
{\it M.\ Vysotsky:} \\
I think that the difference is within the uncertainties of the analyses.
\vspace{5mm} \\
{\it A.L.\ Kataev, INR--Moscow:} \\
What is the main source of theoretical uncertainties in the value of
$\alpha_s(M_Z)$ extracted from fits of the LEP data? Is it sensitive to
the value of the top quark mass? What will happen with the value and
uncertainties of $\alpha_s(m_Z)$ if in fits the top quark mass is fixed?
\vspace{5mm} \\
{\it M.\ Vysotsky:} \\
Theoretical uncertainties in $\alpha_s(m_Z)$ come from higher order terms
which have not been calculated yet.
The sensitivity of $\alpha_s(m_Z)$ to the
value of $m_{top}$ is very small.
\vspace{5mm} \\
{\it M.A.\ Shifman, Minnesota:} \\
If you compare the value of $\alpha_s(m_Z)$ as measured at LEP to that
from low energy data (DIS, jets, {\it etc}.) it comes out to be three
standard deviations higher. If for some unknown reason this LEP measurement
is wrong how would it influence your discussion of $\Gamma(b\bar b)$?
\vspace{5mm} \\
{\it M.\ Vysotsky:} \\
The LEP determination of the $\alpha_s(M_Z)$ value comes mainly from
the ratio $R_l=\Gamma_{hadron}/\Gamma_{lepton}$. I discussed discrepancies
in $R_b=\Gamma_{b\bar b}/\Gamma_{hadron}$; the experimental determination
of $R_b$ does not depend on that of $R_l$. The theoretical expression for
$R_b$ has a very small dependence on $\bar{\alpha}_s$ and its numerical
value practically does not change even if you shift $\alpha_s$ by $3\sigma$.
\vspace{5mm} \\
{\it G.\ Mitselmakher, Fermilab:} \\
How is the Born approximation analysis dependent on the choice of
$\sin\theta$ definition? Also what is the physical reason behind the
choice you use?
\vspace{5mm} \\
{\it M.\ Vysotsky:} \\
The proper choice of $\sin\theta$ is very important in the definition
of the Born approximation. In our choice we take into account
the electromagnetic
coupling running from the fine structure value $\alpha^{-1}
\approx 137$ to its value at the scale of the weak interactions.
\vspace{5mm} \\
{\it D.\ Schildknecht, Bielefeld:} \\
You correctly pointed out that the data require radiative corrections
beyond the $\alpha(m^2_Z)$ Born approximation. You did not comment,
however, on the nature of these corrections. The additional bosonic
corrections have indeed been clearly identified as essentially vertex
corrections which are independent of the mass of the Higgs within the
standard model but are dependent on the empirically unknown trilinear
boson vertex. These corrections are required beyond the full fermion
loops, which contribute not only to the running of $\alpha$ but also
to the $W^\pm$ and $Z^0$ propagators, including light fermions and the
top quark.

I think these beautiful results may be of interest both for theorists
and experimentalists.

\end{document}